\documentclass[english,aps,prl,amsmath,twocolumn,preprintnumber,superscriptaddress,showpacs,notitlepage]{revtex4-1}
\usepackage[T1]{fontenc}
\usepackage[latin9]{inputenc}
\makeatletter
\usepackage{mathptmx,newtxtext,newtxmath,xspace}
\usepackage{amsbsy,bm,bbold}
\usepackage{graphicx,color,xcolor,epsfig,rotate}
\usepackage{fancyhdr}
\usepackage{gensymb}
\usepackage[colorlinks=true, 
            linkcolor=blue, 
            urlcolor=blue,
            citecolor=blue]{hyperref}
\usepackage{soul}
\setstcolor{red}
\newcommand{\iu}{{i\mkern1mu}}

\makeatother

\usepackage{babel}
\begin{document}
\title{Skyrmion Crystal from RKKY Interaction Mediated by 2D Electron Gas}
\author{Zhentao~Wang}
\affiliation{Department of Physics and Astronomy, The University of Tennessee,
Knoxville, Tennessee 37996, USA}
\author{Ying~Su}
\affiliation{Theoretical Division, T-4 and CNLS, Los Alamos National Laboratory,
Los Alamos, New Mexico 87545, USA}
\author{Shi-Zeng~Lin}
\affiliation{Theoretical Division, T-4 and CNLS, Los Alamos National Laboratory,
Los Alamos, New Mexico 87545, USA}
\author{Cristian~D.~Batista}
\affiliation{Department of Physics and Astronomy, The University of Tennessee,
Knoxville, Tennessee 37996, USA}
\affiliation{Quantum Condensed Matter Division and Shull-Wollan Center, Oak Ridge
National Laboratory, Oak Ridge, Tennessee 37831, USA}
\date{\today}
\begin{abstract}
We consider a C$_6$ invariant lattice of magnetic moments coupled via a Kondo exchange $J$ with a 2D electron gas (2DEG). The effective Ruderman-Kittel-Kasuya-Yosida interaction between the moments stabilizes a magnetic skyrmion crystal in the presence of magnetic field and easy-axis anisotropy. An attractive aspect of this mechanism is that the magnitude of the magnetic ordering wave vectors, $\bm{Q}_{\nu}$ ($\nu=1,2,3$), is dictated by the Fermi wave number $k_F$: $|\bm{Q}_{\nu} |=2k_F$. Consequently, the topological contribution to the Hall conductivity of the 2DEG becomes of the order of the quantized value, $e^2/h$, when $J$ is comparable to the Fermi energy $\epsilon_F$.
\end{abstract}
\pacs{~}

\maketitle

The discovery of magnetic skyrmion crystals (SkX) envisioned by Bogdanov and Yablonskii~\citep{Bogdanov89,Rosler2006} in chiral magnets, such as MnSi, Fe$_{1-x}$Co$_x$Si, FeGe and Cu$_2$OSeO$_3$~\citep{Muhlbauer2009, Yu2010a,Yu2011,Seki2012,Adams2012} sparked the interest of the condensed matter community  
at large and spawned efforts in multiple directions. Among those, identifying the basic ingredients for stabilizing SkX in other materials is one of the most pressing challenges  because new stabilization mechanisms are typically accompanied by novel physical properties. For instance, the vector chirality is fixed in the magnetic skyrmions of chiral magnets, such as the so-called B20 compounds; while it is a degree of freedom in the SkX of centrosymmetric materials, such as BaFe$_{1-x-0.05}$Sc$_x$Mg$_{0.05}$O$_{19}$, La$_{2-2x}$Sr$_{1+2x}$Mn$_2$O$_7$, Gd$_2$PdSi$_3$, and Gd$_3$Ru$_4$Al$_{12}$~\cite{Yu2012_BFSMO,Yu2014_biskyrmion,Mallik1998_paramana,Saha1999,Kurumaji2019,Chandragiri_2016,Hirschberger2018}. In the former case, the underlying spiral structure 
is stabilized by a competition between ferromagnetic exchange and the Dzyaloshinskii-Moriya interaction~\cite{Dzyaloshinsky1958,Moriya1960}. In contrast, the spiral ordering of centrosymmetric materials arises from  competition between different exchange couplings or dipolar interactions~\citep{Okubo12,Leonov2015,Lin2016_skyrmion,Hayami16,Batista16}.

To date, most magnetic SkX have been reported in metals, where the interplay between magnetic moments and conduction electrons enables novel response functions, such as the well-known topological Hall effect~\citep{Onoda04,Yi09,Hamamoto15,Gobel2017} and the current-induced skyrmion motion~\citep{Jonietz2010,Yu2012,Schulz2012,Nagaosa2013}. The topological Hall effect is a direct consequence of the Berry curvature acquired by the reconstructed electronic bands. In the adiabatic limit, the momentum space Berry curvature is controlled by a real space Berry curvature that is proportional to the skyrmion density: each skyrmion produces an effective flux equal to the flux quantum $\Phi_0$. Consequently, Hall conductivities comparable to the quantized value ($e^2/h$)  can in principle be achieved if the ordering wave vector of the SkX is comparable to the Fermi wave vector $k_F$. As we demonstrate in this Letter, this condition can be naturally fulfilled in $f$-electron systems where the Ruderman-Kittel-Kasuya-Yosida (RKKY) interaction is mediated by conduction electrons~\citep{Ruderman,Kasuya,Yosida1957}. 
Our results are potentially relevant for the rare earth based materials Gd$_2$PdSi$_3$ and Gd$_3$Ru$_4$Al$_{12}$ that contain a magnetic field induced SkX phase in their phase diagrams~\cite{Mallik1998_paramana,Saha1999,Kurumaji2019,Chandragiri_2016,Hirschberger2018}.

We first demonstrate that the magnetic susceptibility of a 2D electron gas (2DEG) on a C$_6$ invariant lattice has a maximum at $2k_F$ whenever the single-electron dispersion relation,
\begin{equation}
\epsilon_{\bm{k}}\approx\frac{1}{2m}\left(k^{2}+u k^{4}\right),\label{eq:k2k4}
\end{equation}
has a negative quartic correction $u\equiv w/k_{F}^{2} <0$. Under this condition, a small easy-axis anisotropy is enough to stabilize a magnetic-field induced SkX, which is approximately described by the superposition of three spirals with ordering wave vectors $\bm{Q}_{\nu}$ ($\nu=1,2,3$), that are related by $\pm 2\pi/3$ rotations. Given that 
$| {\bm Q}_{\nu}| = 2k_F$, the resulting SkX produce a very large Hall conductivity (of order $e^2/h$) for a Kondo exchange of $J/\epsilon_F \approx 0.3$. This condition can only be fulfilled in the dilute limit $\epsilon_F \eta(\epsilon_{F})\approx (k_F^2 a^2/ 2 \pi) \ll 1$, where $\eta(\epsilon)\approx m/(2\pi)$ is the density of states and $a$ is the lattice constant. Because we are interested in the regime of weak Kondo effect, here we only consider the classical limit of the local magnetic moments.

We start by considering the 2D Kondo lattice model (KLM) for classical magnetic moments:
\begin{equation}
\mathcal{H}=\sum_{\bm{k}}\sum_{\sigma}\left(\epsilon_{\bm{k}}-\mu\right)c_{\bm{k}\sigma}^{\dagger}c_{\bm{k}\sigma}+J\sum_{i}\sum_{\alpha\beta}c_{i\alpha}^{\dagger}\bm{\sigma}_{\alpha\beta}c_{i\beta}\cdot\bm{S}_{i},\label{eq:Kondo}
\end{equation}
where the operator $c_{i\sigma}^{\dagger}$ ($c_{i\sigma}$) creates (annihilates)
an itinerant electron on site $i$ with spin $\sigma$, and $c_{\bm{k}\sigma}^{\dagger}$
($c_{\bm{k}\sigma}$) is the corresponding Fourier transform. $\epsilon_{\bm{k}}$
is the bare electron dispersion with chemical potential $\mu$. $J$
is the exchange interaction between the local magnetic moments $\bm{S}_{i}$
and the conduction electrons ($\bm{\sigma}$ is the vector of the
Pauli matrices) and $\left|\bm{S}_{i}\right|=1$.

In the weak-coupling limit, $J\eta(\epsilon_{F})\ll1$,
the interaction between local moments is described by the RKKY model:
\begin{equation}
\mathcal{H}_{\text{RKKY}}=-J^{2}\sum_{\bm{k}}\chi_{\bm{k}}\bm{S}_{\bm{k}}\cdot\bm{S}_{-\bm{k}},\label{eq:RKKY}
\end{equation}
with
\begin{align}
\chi_{\bm{k}} & =-\frac{2}{V_{\text{BZ}}}\int d\bm{q}\frac{f(\epsilon_{\bm{q}+\bm{k}})-f(\epsilon_{\bm{q}})}{\epsilon_{\bm{q}+\bm{k}}-\epsilon_{\bm{q}}},\label{eq:chi0}\\
\bm{S}_{\bm{k}} & =\frac{1}{\sqrt{N}}\sum_{i}e^{\iu\bm{k}\cdot\bm{r}_{i}}\bm{S}_{i},\label{eq:Fourier_S}
\end{align}
where $N$ is the number of lattice sites, and $f(\epsilon)$
is the Fermi distribution function.

\begin{figure}
\centering
\includegraphics[width=1\columnwidth]{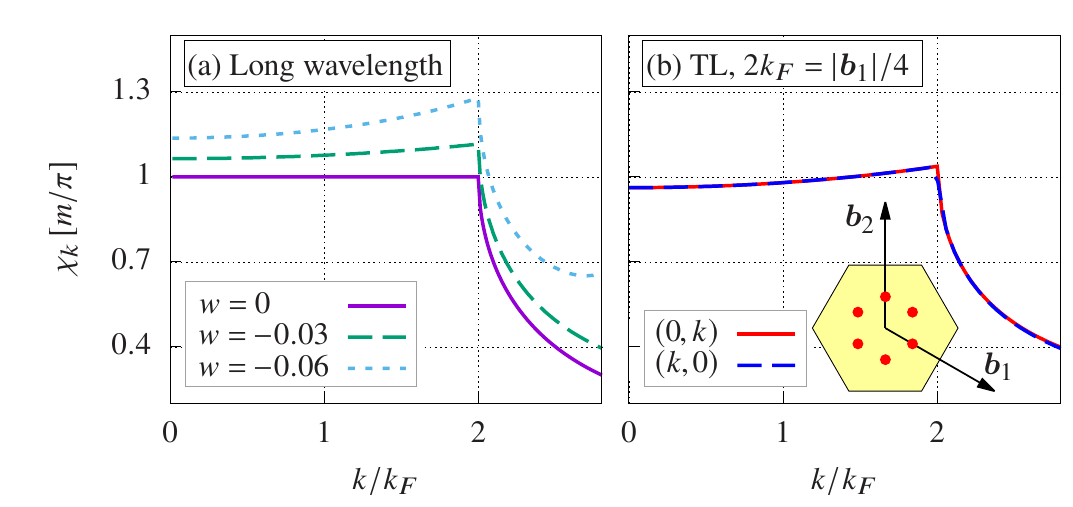}
\caption{Susceptibility $\chi_{\bm{k}}$ for different dispersions. (a) Long
wavelength limit Eq.~\eqref{eq:k2k4}, using $V_\text{BZ}=(2\pi)^2$; (b) TL Eq.~\eqref{eq:dispersion_TL}
with $2k_{F}=|\bm{b}_{1}|/4$. The inset shows the peak positions
of $\chi_{\bm{k}}$ of the TL in the first BZ, where $\{\bm{b}_{1}\,,\bm{b}_{2}\}$
are the reciprocal space basis vectors. \label{fig:chi}}
\end{figure}

In general, the RKKY interaction depends on the details of the Fermi surface (FS).
However, for small filling fraction, the electronic dispersion relation of C$_6$ invariant systems can be approximated by 
Eq.~\eqref{eq:k2k4} and the FS is circular. In absence of the 
quartic term $(w=0)$, the susceptibility is
flat below $2k_{F}$~\citep{Giuliani_book} {[}see Fig.~\ref{fig:chi}(a){]},
\begin{equation}
\chi_k=\frac{m}{\pi}\left[1-\Theta\left(p-2\right)\sqrt{1-\left(2/p\right)^{2}}\right],
\end{equation}
with $p\equiv k/k_F$. The discontinuity of $\partial_k \chi_k$ at $k=2 k_F$ is related to the long-range nature of the RKKY interaction (it decays as $1/r^2$ in real space). The resulting  RKKY model is highly frustrated because any spiral ordering with wave number $q \leq 2k_{F}$ is a ground state. 

The frustration is partially lifted by a finite quartic term and the magnetic susceptibility becomes
\begin{equation}
\begin{split}\chi_k & =\frac{m}{\pi\left|w\right|p^{2}\lambda(p,w)}\Bigg[2\arctan\frac{1}{\lambda(p,w)} \\
&\quad -2\Theta(p-2)\arctan\frac{\sqrt{1-(2/p)^{2}}}{\lambda(p,w)}\\
&\quad+\arctan\frac{\left|1+wp^{2}\right|}{\left|w\right|p^{2}\lambda(p,w)} \\
&\quad -\arctan\frac{\sqrt{(1+wp^{2})^{2}+4w(w+1)}}{\left|w\right|p^{2}\lambda(p,w)}\Bigg],
\end{split}
\label{eq:chi_with_w}
\end{equation}
for $-1\ll w < 0$ and $p^2<2/|w|$, where $\lambda(p,w)\equiv\sqrt{\left|1+2/(wp^{2})\right|}$.
As shown in Fig.~\ref{fig:chi}(a), $\chi_k$ is maximized on the ring $k=2k_{F}$,  where the function $\chi_k$ 
is nonanalytical. 
As we will see below, the control parameter for the stability of the SkX is the ratio $\chi_{2k_F}/\chi_{k=0}$. According to Eq.~\eqref{eq:chi_with_w},  $\chi_{2k_F}/\chi_{k=0}$ is determined by $w \equiv u k_F^2$ for $k_F \ll 1$, i.e., $w$ becomes the control parameter in the long wave length regime.


For concreteness, we consider a triangular lattice (TL) with hopping
amplitudes $\{t,\,t_{2},\,t_{3}\}$ for the first, second, and third nearest
neighbors (from now on, we will set $t_{2}=t_{3}=0$ unless specified otherwise):
\begin{align}
\epsilon_{\bm{k}}&= -2t\left[\cos k_{x}+2\cos\frac{k_{x}}{2}\cos\frac{\sqrt{3}k_{y}}{2}\right]\nonumber \\
 & \quad-2t_{2}\left[\cos\left(\sqrt{3}k_{y}\right)+2\cos\left(\frac{3k_{x}}{2}\right)\cos\left(\frac{\sqrt{3}k_{y}}{2}\right)\right]\nonumber \\
 & \quad-2t_{3}\left[\cos\left(2k_{x}\right)+2\cos k_{x}\cos\left(\sqrt{3}k_{y}\right)\right].\label{eq:dispersion_TL}
\end{align}
The mass $m$, and $w$ are obtained by expanding $\epsilon_{\bm{k}}$ near $k=0$:
\begin{equation}
m=\frac{1}{3\left(t+3t_{2}+4t_{3}\right)},\quad w=-\frac{k_{F}^{2}}{16}\frac{t+9t_{2}+16t_{3}}{t+3t_{2}+4t_{3}}.\label{eq:mw}
\end{equation}

The resulting magnetic susceptibility, shown in Fig.~\ref{fig:chi}(b), confirms that
$\chi_{\bm{k}}$ is maximized on the $k=2k_{F}$ ring. 
The degeneracy of the ordering wave vectors along the ring direction is lifted by lattice anisotropy terms of order $\mathcal{O}(k^{6})$:
while the angular dependence of $\chi_{\bm{k}}$ is very small, 
a careful numerical integration of Eq.~\eqref{eq:chi0}
shows only six discrete peaks ($\pm\bm{Q}_{\nu=1,2,3}$) with the
same amplitude {[}see inset of Fig.~\ref{fig:chi}(b){]}.

An immediate question is what magnetic
structure is stabilized in the presence of magnetic field and easy-axis anisotropy~\footnote{The easy-axis anisotropy is generated  by the combined effect of the spin-orbit coupling and the crystal field on the magnetic ion}. 
It has been shown that SkX can arise in hexagonal frustrated Mott insulators whose exchange interactions lead to a similar set of six maxima in the magnetic susceptibility~\cite{Bogdanov89,Okubo12,Leonov2015,Hayami16}. 
Indeed, the phase diagram of these materials can be described with a generic Ginzburg-Landau (GL) theory, which only assumes that
the magnetic susceptibility is maximized over a ring of wave vectors of the same magnitude and that it is an analytic function of $k$~\cite{Lin2016_skyrmion}. The nonanalytical behavior of $\chi_k$ at $k=2k_F$ violates the second assumption, and raises the question of whether the SkX phase can still be stabilized in RKKY systems.
Motivated by this question, we  add the corresponding Zeeman and anisotropy terms to the TL RKKY Hamiltonian:
\begin{equation}
\mathcal{H}_{\text{total}}=\mathcal{H}_{\text{RKKY}}+\mathcal{H}^{\prime},\quad\mathcal{H}^{\prime}=-H\sum_{i}S_{i}^{z}+D\sum_{i}\left(S_{i}^{z}\right)^{2}.\label{eq:H_total}
\end{equation}

Our Monte Carlo simulation with Metropolis update on finite
lattices indeed suggests that the ordering wave number coincides with
 $2k_F$ at low temperature. However, due to the highly
frustrated nature of the RKKY model, the Metropolis update is not
efficient enough to overcome freezing into metastable states. 
We then adopt a $T=0$ variational approach, which further confirms that the  magnetic ordering wave vector has magnitude $Q=2k_F$~\cite{supp}.

\begin{figure}
\centering
\includegraphics[width=1\columnwidth]{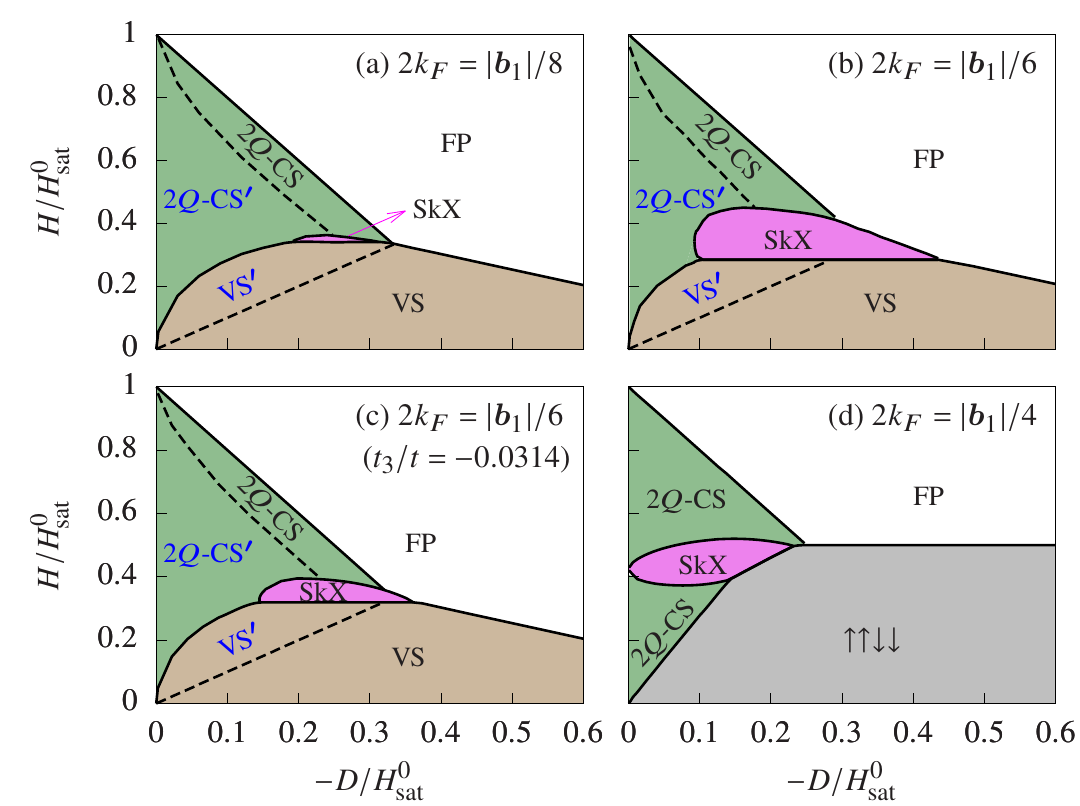}
\caption{Phase diagrams of the TL RKKY model with easy-axis single-ion anisotropy
in a magnetic field. We set $t_{2}=t_{3}=0$ except in (c). 
(a) $2k_F=|\bm{b}_1|/8$, which gives $\chi_{2k_F}/\chi_{k=0}\approx 1.0177$, $w\approx-0.013$, $H_{\text{sat}}^{0}\approx0.0034J^{2}/t$; 
(b) $2k_F=|\bm{b}_1|/6$, which gives $\chi_{2k_F}/\chi_{k=0}\approx 1.0323$, $w\approx-0.023$, $H_{\text{sat}}^{0}\approx0.0062J^{2}/t$; 
(c) $2k_F=|\bm{b}_1|/6$, which gives $\chi_{2k_F}/\chi_{k=0}\approx 1.0211$, $w\approx-0.013$, $H_{\text{sat}}^{0}\approx0.0046J^{2}/t$; 
(d) $2k_F=|\bm{b}_1|/4$, which gives $\chi_{2k_F}/\chi_{k=0}\approx 1.0783$, $w\approx-0.051$, $H_{\text{sat}}^{0}\approx0.016J^{2}/t$.
\label{fig:phd}}
\end{figure}

\begin{figure}
\centering
\includegraphics[width=1\columnwidth]{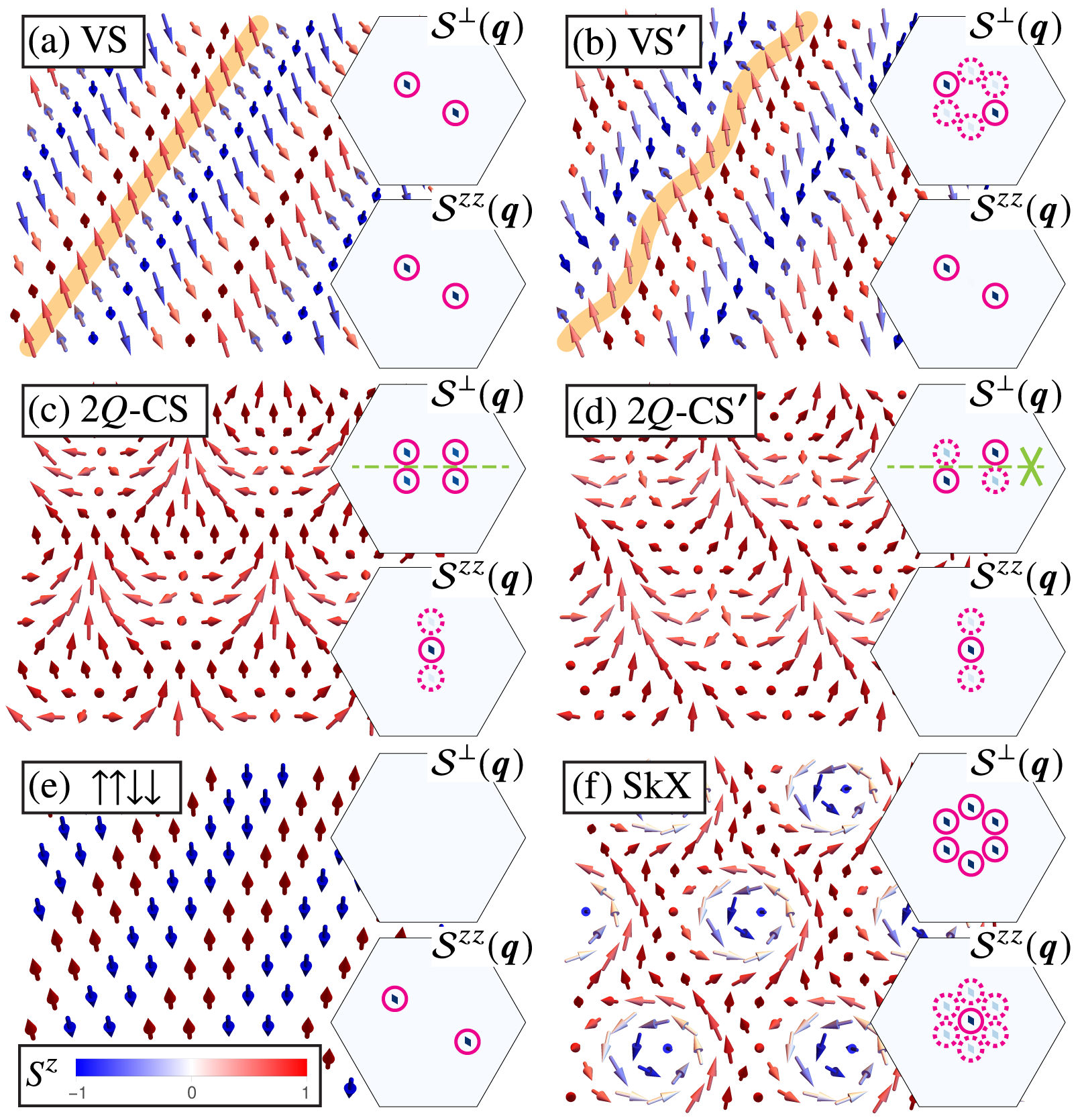}
\caption{Spin configurations of phases shown in Fig.~\ref{fig:phd}. The insets show the in-plane ($\mathcal{S}^{\perp}$)
and out-of-plane ($\mathcal{S}^{zz}$) static structure factors in
the first BZ. The solid (dotted) circles highlight the dominant (subdominant)
peaks.
The orange markups in (a),(b) highlight the difference between VS and VS$^{\bm{\prime}}$, i.e., absence or presence of the in-plane modulation. The green markups in insets of (c),(d) show the difference between $2\bm{Q}$-CS and $2\bm{Q}$-CS$^{\bm{\prime}}$: the former is invariant under a mirror reflection of the $xz$ plane followed by a $\pi$ rotation along the $x$ axis, while the latter does not respect this symmetry.
\label{fig:spin}}
\end{figure}

Figure~\ref{fig:phd} shows the $T=0$ phase diagrams for 
$2k_F=\{|\bm{b}_1|/8,\,|\bm{b}_1|/6,\,|\bm{b}_1|/4 \}$
including seven different phases,
namely, the vertical spiral (VS), vertical spiral with in-plane modulation
(VS$^{\bm{\prime}}$), $2\bm{Q}$-conical spiral ($2\bm{Q}$-CS), $2\bm{Q}$-conical
spiral with unequal in-plane structure factor intensities ($2\bm{Q}$-CS$^{\bm{\prime}}$),
up-up-down-down ($\uparrow\uparrow\downarrow\downarrow$), SkX, and the fully polarized (FP) phases (see Fig.~\ref{fig:spin}).

The phase diagrams shown in Fig.~\ref{fig:phd} are similar to the one obtained from short-range
Heisenberg models on the TL~\citep{Leonov2015}, and more generally, from a GL
analysis of the inversion-symmetric magnets~\citep{Lin2016_skyrmion}. 
However, there is  an important qualitative difference associated with the stability of the SkX phase.

A direct comparison between Figs.~\ref{fig:phd}(a) and \ref{fig:phd}(b) suggests that the size
of the SkX phase depends sensitively on the ratio $\chi_{2k_F}/\chi_{k=0}$,
which is controlled by the parameter $w$ from the long wavelength
analysis {[}see Eq.~\eqref{eq:chi_with_w}{]}. To reveal the role
of $w$, we keep $2k_{F}=|\bm{b}_{1}|/6$ and add a finite
$t_{3}$, that changes $w$ from $-0.023$ to $-0.013$ {[}see Fig.~\ref{fig:phd}(c){]}.
The SkX phase shrinks as we decrease $|w|$ while keeping
$2k_{F}$ unchanged {[}Figs.~\ref{fig:phd}(b) and \ref{fig:phd}(c){]}. 
Indeed, our variational approach confirms
that the 
SkX phase disappears for $|w|\lesssim 0.0115$ when using $\chi_{\bm{k}}$ given by Eq.~\eqref{eq:chi_with_w} (valid for $k_F \ll 1$).
In other words, the SkX phase is stable  in the {\it mesoscale} regime $k_F > \sqrt{0.0115/|u|}$.
 
This behavior is qualitatively different from the GL theory, where the phase diagram, including the SkX phase, remains invariant upon approaching the Lifshitz transition~\cite{Lin2016_skyrmion}. 
The key difference is in the relative difference between $(\chi_{2k_F}-\chi_{k=0})$ that determines the saturation field $H_{\text{sat}}^{0}\equiv2J^{2}\left(\chi_{2k_{F}}-\chi_{k=0}\right)$, and  $(\chi_{2k_F}-\chi_{{\bm k}_{n}})$ for
${\bm k}_{n} = \sum_{\nu=1,3} n_{\nu} {\bm Q}_{\nu}$ ($n_{\nu}$ are {\it small} integer numbers and $|\bm{k}_n| >Q$) that determines the exchange energy of the higher harmonics  present in most of the phases, including the SkX. 
In the GL theory, both energy scales remain comparable for $Q \to 0$. In contrast, $(\chi_{2k_F}-\chi_{{\bm k}_{n}})$ becomes much bigger than 
$(\chi_{2k_F}-\chi_{k=0})$ for $k_F \to 0$ because the slope of $\chi_k$ diverges for $k \to 2 k_F^{+}$. This difference becomes less important for $\sqrt{0.0115/|u|} < k_F < 1$, explaining why the phase diagrams of both theories  become very similar in the mesoscale regime.

Figures~\ref{fig:phd}(a) and \ref{fig:phd}(c) show the phase diagrams  obtained for the same value of
$w$ and different values of $2k_{F}$ ($2k_{F}=|\bm{b}_{1}|/6$ and $2k_{F}=|\bm{b}_{1}|/8$). The difference between both phase diagrams is produced by small deviations  of $\chi_k$ from the universal expression  in Eq.~\eqref{eq:chi_with_w}. 
The size of the SkX phase is bigger for $2k_{F}=|\bm{b}_{1}|/6$ simply because  $\chi_{2k_F}/\chi_{k=0}$ is bigger.

For large enough $2k_{F}$, the long wavelength
analysis is no longer accurate and the phase diagram can become qualitatively different from the one obtained for small ordering wave numbers.
This is illustrated by Fig.~\ref{fig:phd}(d) for $2k_F=|{\bm b}_1|/4$:
the low field VS and VS$^{\bm{\prime}}$ phases are
replaced with a collinear $\uparrow\uparrow\downarrow\downarrow$ ordering, and the low field $2\bm{Q}$-CS$^{\bm{\prime}}$ phase is replaced by the $2\bm{Q}$-CS phase. Interestingly, we find
that the SkX phase remains robust.

\begin{figure}
\centering
\includegraphics[width=1\columnwidth]{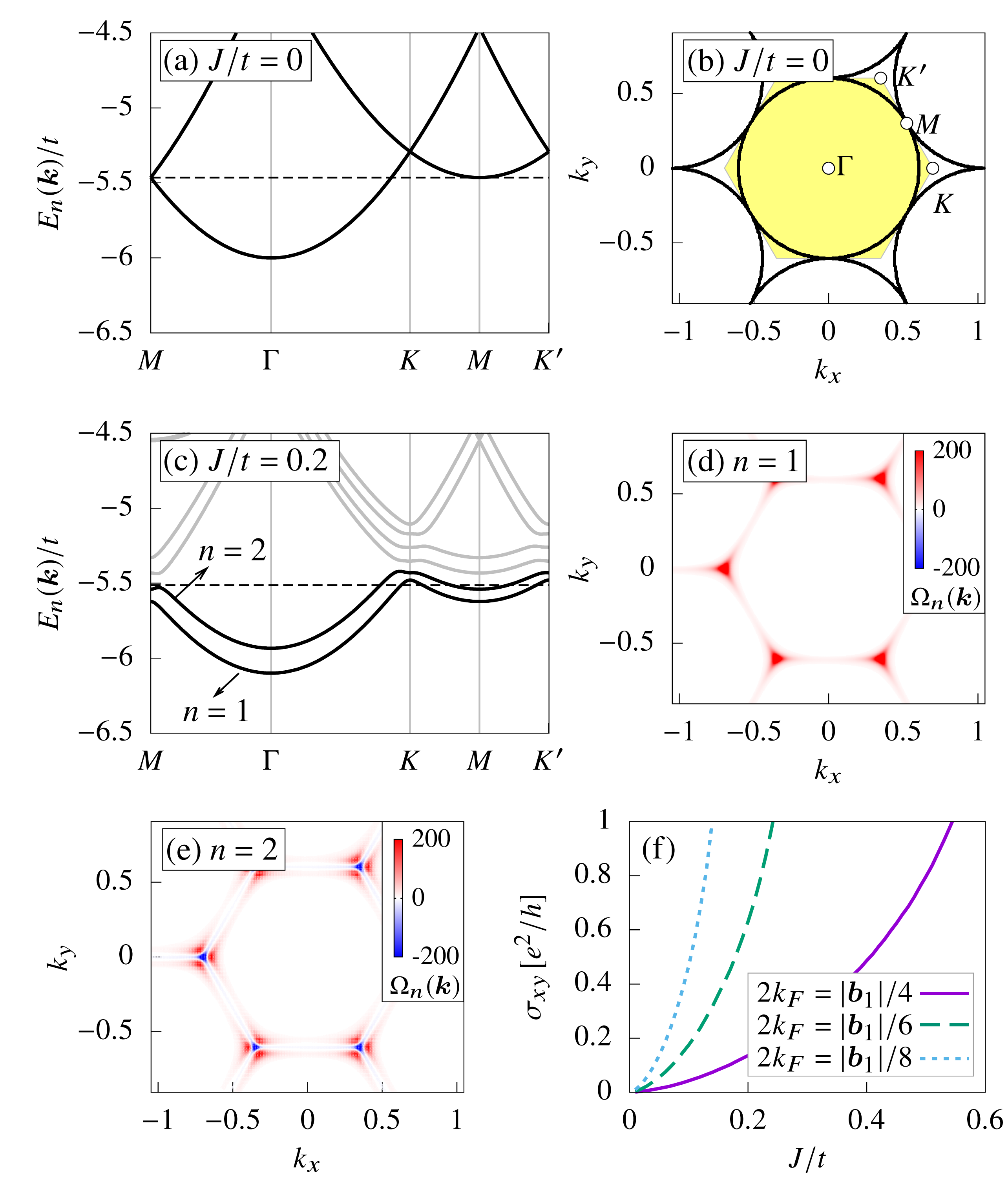}
\caption{(a),(b) Electronic band structure and FS of the TL KLM in
the folded first BZ ($2k_F=|\bm{b}_{1}|/6$) with $J=0$. (c)--(e) Electronic
band structure and Berry curvatures of the lowest two bands in the
folded first BZ ($2k_F=|\bm{b}_{1}|/6$) with $J/t=0.2$ and SkX spin configuration.
(f) Transverse conductivities of the TL KLM with SkX spin configurations
of different $Q$, at fixed electron filling fractions. The horizontal
dashed lines in (a),(c) show the Fermi level for fixed filling $n_{c}\approx0.0252$.
We use SkX spin configuration obtained at \{$D=-0.002J^{2}/t$, $H=0.007J^{2}/t$\}
for $2k_F=|\bm{b}_{1}|/4$, \{$D=-0.0015J^{2}/t$, $H=0.002J^{2}/t$\}
for $2k_F=|\bm{b}_{1}|/6$, and \{$D=-0.0008J^{2}/t$, $H=0.00117J^{2}/t$\}
for $2k_F=|\bm{b}_{1}|/8$. \label{fig:berry}}
\end{figure}

The SkX phase induces nontrivial Berry curvature and anomalous Hall
response when coupled to itinerant electrons~\cite{Onoda04,Yi09,Hamamoto15,Gobel2017}. 
Figures~\ref{fig:berry}(a) and \ref{fig:berry}(b) show the folded unreconstructed electronic band structure ($J=0$) 
and the FS in the reduced BZ ($2k_F=|\bm{b}_{1}|/6$).
A finite coupling $J$ opens a gap at the $M$ and $K$ points {[}Fig.~\ref{fig:berry}(c){]}.
The lowest two bands develop nonzero Berry curvature $\Omega_{n}(\bm{k})$
centered at $K$ and $K^{\prime}$, where the electron wave functions
have the largest renormalization [Figs.~\ref{fig:berry}(d) and \ref{fig:berry}(e)]. 
We note that the two lowest bands
have the same Chern number: both $C_{n}=1$ or $C_{n}=-1$,
where the sign is determined by the sign of the scalar spin chirality.

Figure~\ref{fig:berry}(f) shows the transverse conductivity
$\sigma_{xy}$~\cite{Xiao2010_RMP} of the TL KLM in the SkX phase with fixed electron
fillings. A large $\sigma_{xy}$ can be achieved even
in the weak-coupling regime ($J/t\ll1$), and its magnitude increases
quickly upon approaching the long wavelength limit.
Such a behavior is {\it opposite} to the conventional understanding based on the strong-coupling or adiabatic limit. The skyrmion density is no longer dictated by $k_F$ in that limit and it is typically much smaller than the electron density. Thus, a larger skyrmion density (bigger $Q$) produces larger effective magnetic field and consequently a larger Hall response in the strong-coupling limit.

The $k_{F}$-dependence of $\sigma_{xy}$ can be understood in a simple way.
In Fig.~\ref{fig:berry}(b), we see that the FS already occupies
$91\%$ of the folded first BZ. 
$\sigma_{xy}$ becomes large (of order $e^2/h$) when the states near
$K$ and $K^{\prime}$ points (where most of the Berry curvature concentrates) are pushed below the Fermi level.
This condition is fulfilled when the gap, of order $J$, becomes comparable to the energy difference between $\epsilon_K$ and $\epsilon_{k_F}$:
\begin{equation}
J\sim\frac{k_{F}}{m}\left(\frac{2}{\sqrt{3}}k_{F}-k_{F}\right)\approx0.155\frac{k_{F}^{2}}{m} \approx 0.3 \epsilon_F.
\end{equation}
In other words, in agreement with the result shown in Fig.~\ref{fig:berry}(f), the value of $J$ required to produce a large anomalous Hall effect is smaller for smaller 
values of $k_{F}$. In fact, the three curves shown in Fig.~\ref{fig:berry}(f) collapse into a single curve
after rescaling the $x$ axis to $J/\left(tk_{F}^{2}\right)$.

To summarize, we find that the quartic term in the dispersion of
a 2DEG induces susceptibility maxima at $Q=2k_{F}$,
which is key to produce a finite saturation field and to stabilize the spiral ordering 
that is the basis for generating SkX via RKKY interactions. 
The $T=0$ phase diagram includes a sizable SkX
phase induced by easy-axis anisotropy and a magnetic field. 
The size of the SkX phase and its stability range 
is controlled
by $\chi_{2k_F}/\chi_{k=0}$, which is in turn determined by $w$ for $k_F \ll 1$. 
The same electrons which
induce the RKKY interaction and the SkX phase, exhibit a large anomalous
Hall {\it counter-response}, whose magnitude depends solely on $J/\left(tk_{F}^{2}\right)$
in the weak-coupling limit. 
This strong feedback effect distinguishes the stabilization mechanism based 
on the RKKY interaction from other mechanisms in which the lattice parameter of the SkX and the Fermi wave-length are independent length scales.

We note that the $\mathcal{O}(k^4)$ correction in the quasi-particle dispersion relation \eqref{eq:k2k4} is not the only way of lifting the spin susceptibility degeneracy of the 2DEG for $k < 2 k_F$.
For example, the electron-electron interactions can also induce a global maximum of $\chi_{\bm{k}}$ at a finite wave number close to $2k_F$~\cite{Belitz97,Hirashima98,Chubukov03,Gangadharaiah05,Chubukov05,Maslov06,Schwiete06,Aleiner06,Shekhter06,Simon2008,Prus03}. 
Additional magnetic interactions, such as short-range superexchange and dipolar coupling can produce a similar effect.
It is also important to note, that four-spin and higher order interactions, not included in the RKKY Hamiltonian, are naturally generated from the KLM upon moving away from the weak-coupling limit~\cite{Kumar2010,Akagi2012,Solenov2012,Hayami2014,Ozawa2016,Batista16,Ozawa2017,Hayami2017}. 
These higher order terms can also stabilize multi-$\bm{Q}$ magnetic 
orderings that include SkX phases~\cite{Solenov2012,Hayami2014,Ozawa2016,Batista16,Ozawa2017,Hayami2017}. 
A variational treatment like the one that has been presented  in this Letter can be applied to the full KLM to determine if the effective higher order spin interactions can further stabilize the field-induced SkX phase.

Our results are potentially relevant for explaining the giant topological Hall response of  Gd$_2$PdSi$_3$~\cite{Mallik1998_paramana,Saha1999}, which is produced by a field induced SkX phase, as it has been recently revealed by resonant x-ray scattering~\cite{Frontzek_thesis2009,Kurumaji2019}. 
Indeed, angle-resolved photoemission spectroscopy suggests that 
RKKY is the dominant interaction in this material~\cite{Inosov2009}. 
More generally, SkX phases induced by RKKY interactions are expected to be realized in a wider range of hexagonal intermetallic compounds with in-plane spiral ordering (ordering wave vector parallel to the plane) and moderate easy-axis anisotropy.

We thank K.~Barros and H.~Ishizuka for helpful discussions. 
Z.\nobreak\,W. and C.\nobreak\,D.\nobreak\,B. are supported by funding from the Lincoln Chair of Excellence in Physics.
The work at Los Alamos National Laboratory (LANL) was carried out
under the auspices of the U.S. DOE NNSA under Contract No.
89233218CNA000001 through the LDRD Program, and was supported by the
Center for Nonlinear Studies at LANL.
This research used resources of the Oak Ridge Leadership Computing
Facility at the Oak Ridge National Laboratory, which is supported
by the Office of Science of the U.S. Department of Energy under Contract
No. DE-AC05-00OR22725. 

\bibliographystyle{apsrev4-1}
\bibliography{ref}

\setcounter{figure}{0} 
\renewcommand{\thefigure}{S\arabic{figure}} 
\setcounter{equation}{0} 
\renewcommand{\theequation}{S\arabic{equation}}

\begin{center}   
{\bf ---Supplemental Material---} 
\end{center}

\section{Variational Method}

Here we describe the variational method used to produce the $T=0$ phase diagrams in the main text. 
We will start by assuming that the ground state ordering wavelength $Q$ is fixed by $2k_F$. This constraint will be removed later.
In this case, the magnetic super-cell contains $L\times L$ spins spanned by the basis \{$L\bm{a}_1$, $L\bm{a}_2$\}, where $\bm{a}_1$ and $\bm{a}_2$ are the basis of the triangular lattice.
Here, we use $Q=2k_F = l |\bm{b}_1|/L$, and \{$l$, $L$\} are the minimum integers satisfying the equation.
The choice of commensurate $2k_F$ is only for the convenience of calculation, and any nearby incommensurate $2k_F$ should produce similar phase diagrams.

The spin states can be described by $3\times L \times L$ variational parameters \{$S_{\bm{r}}^x, \, S_{\bm{r}}^y, \,  S_{\bm{r}}^z $\} and $L\times L$ equality constraints $|\bm{S}_{\bm{r}}|=1$. The total energy density of the model is then minimized in this constrained variational space~\cite{nlopt,Rowan_thesis1990}.

Note that we have not made any assumption about the spin structure, except for the size of the underlying magnetic super-cell size. Clearly, since there are $3\times L \times L$ variational parameters and $L \times L$ equality constraints, the energy minimization is relatively time-consuming. The reward for this effort is that the results are not biased towards any given ansatz of spin configurations.

After obtaining the phase diagrams from the above-mentioned approach, we can perform a Fourier analysis of each phase and parametrize the corresponding spin configurations by a few dominant Fourier components at $\bm{k}={0}$ and $\bm{Q}_\nu$, where we have defined
\begin{equation}
\bm{Q}_1 = q \bm{b}_1,\quad \bm{Q}_2 = q \bm{b}_2, \quad \bm{Q}_3 = -\bm{Q}_1 - \bm{Q}_2,
\end{equation}
and $0<q\le 1/2$.
The results of the Fourier analysis are very similar to those described by Refs.~\cite{Leonov2015,Lin2016_skyrmion}, which we document below.

In the VS and VS$^{\bm{\prime}}$ phases, the normalized spin configurations $\bm{S}_{\bm{r}}\equiv  \bm{m}_{\bm{r}} / | \bm{m}_{\bm{r}} |$ can be parametrized as:
\begin{subequations}\label{eq:ansatz_VS}
\begin{align}
m_{\bm{r}-\bm{r}_0}^x &= a_1 \cos(\phi) \sin(\bm{Q}_1 \cdot \bm{r}) \nonumber \\
&\quad + a_2 \sin(\phi) \left[ \cos(\bm{Q}_2 \cdot \bm{r}) + \cos( \bm{Q}_3 \cdot \bm{r}) \right],\\
m_{\bm{r}-\bm{r}_0}^y &= a_1 \sin(\phi) \sin(\bm{Q}_1 \cdot \bm{r}) \nonumber \\
&\quad - a_2 \cos(\phi) \left[ \cos(\bm{Q}_2 \cdot \bm{r}) + \cos( \bm{Q}_3 \cdot \bm{r}) \right],\\
m_{\bm{r}-\bm{r}_0}^z &= a_0 - a_1 \cos (\bm{Q}_1 \cdot \bm{r}),
\end{align}
\end{subequations}
where the parameter $\phi$ controls the direction of the spiral plane. For the RKKY model considered in this Letter, $\phi$ is an arbitrary number since it does not affect the energy produced by isotropic spin exchanges. The difference between the two phases comes from the parameter $a_2$: it is zero in the VS phase, while it is finite for the VS$^{\bm{\prime}}$ phase.

In the $2\bm{Q}$-CS and $2\bm{Q}$-CS$^{\bm{\prime}}$ phases, the normalized spin configurations $\bm{S}_{\bm{r}}\equiv  \bm{m}_{\bm{r}} / | \bm{m}_{\bm{r}} |$ can be parametrized as:
\begin{subequations}\label{eq:ansatz_CS}
\begin{align}
m_{\bm{r}-\bm{r}_0}^x &= a_1 \sin(\bm{Q}_1 \cdot \bm{r}) + a_2 \sin(\bm{Q}_3 \cdot \bm{r}),\\
m_{\bm{r}-\bm{r}_0}^y &= a_1 \cos(\bm{Q}_1 \cdot \bm{r}) - a_2 \cos(\bm{Q}_3 \cdot \bm{r}), \\
m_{\bm{r}-\bm{r}_0}^z &= a_0 + a_3 \cos(\bm{Q}_2 \cdot \bm{r}).
\end{align}
\end{subequations}
The relation between $a_1$ and $a_2$ distinguishes the two phases: in the $2\bm{Q}$-CS phase we have $a_1 = a_2$, while in $2\bm{Q}$-CS$^{\bm{\prime}}$ phase we have $a_1 \neq a_2$. 

In the SkX phase, the normalized spin configurations $\bm{S}_{\bm{r}}\equiv  \bm{m}_{\bm{r}} / | \bm{m}_{\bm{r}} |$ can be parametrized as:
\begin{widetext}
\begin{subequations}\label{eq:ansatz_SkX}
\begin{align}
m_{\bm{r}-\bm{r}_0}^x &= a_1 \left[ \cos (\varphi) \sin (\bm{Q}_1 \cdot \bm{r} + \theta_1) 
+ \cos \left( \varphi + \kappa \frac{2\pi}{3} \right) \sin (\bm{Q}_2 \cdot \bm{r} + \theta_1)
+ \cos \left( \varphi + \kappa \frac{4\pi}{3} \right) \sin (\bm{Q}_3 \cdot \bm{r} + \theta_1) \right], \\
m_{\bm{r}-\bm{r}_0}^y &= a_1 \left[ \sin (\varphi) \sin (\bm{Q}_1 \cdot \bm{r} + \theta_1) 
+ \sin \left( \varphi + \kappa \frac{2\pi}{3} \right) \sin (\bm{Q}_2 \cdot \bm{r} + \theta_1)
+ \sin \left( \varphi + \kappa \frac{4\pi}{3} \right) \sin (\bm{Q}_3 \cdot \bm{r} + \theta_1) \right], \\
m_{\bm{r}-\bm{r}_0}^z &= a_0 - a_2 \left[ \cos(\bm{Q}_1 \cdot \bm{r} + \theta_2)+\cos(\bm{Q}_2 \cdot \bm{r} + \theta_2)+\cos(\bm{Q}_3 \cdot \bm{r} + \theta_2) \right],
\end{align}
\end{subequations}
\end{widetext}
where the parameter $\kappa=\pm 1$ controls the sign of the scalar chirality, and $\varphi$ controls the helicity of the skyrmion configuration. As we discussed in the main text, both $\kappa$ and $\varphi$ can change freely for the isotropic RKKY model under consideration.

In principle, the wave number $Q$ depends on the magnetic field and easy-axis anisotropy, as it is demonstrated in Ref.~\cite{Bogdanov1994} for metastable SkX in helimagnets.
Within the variational manifold generated by the above-mentioned parametrizations, we can relax the magnetic super-cell assumption, and make $Q$ a variational parameter to check this point. 
More specifically, for each point in the phase diagrams obtained by the previous method (which fixes $Q=2k_F$), we can compare the optimized energies of different ansatze and find the lowest one:
\begin{itemize}
\item The VS and VS$^{\bm{\prime}}$ phases: we minimize the total energy density in the variational space given by Eq.~\eqref{eq:ansatz_VS}. The variational parameters are \{$q$, $a_0$, $a_1$, $a_2$, $\bm{r}_0$\}.
\item The $2\bm{Q}$-CS and $2\bm{Q}$-CS$^{\bm{\prime}}$ phases: we minimize the total energy density in the variational space given by Eq.~\eqref{eq:ansatz_CS}. The variational parameters are \{$q$, $a_0$, $a_1$, $a_2$, $a_3$, $\bm{r}_0$\}.
\item The SkX phase: we minimize the total energy density in the variational space given by Eq.~\eqref{eq:ansatz_SkX}. The variational parameters are \{$q$, $a_0$, $a_1$, $a_2$, $\theta_1$, $\theta_2$, $\bm{r}_0$\}.
\item The FP phase: the total energy density is given by $-J^2 \chi_{k=0} -H + D$.
\item The $\uparrow\uparrow\downarrow\downarrow$ phase: the total energy density is given by $-J^2 \chi_{2k_F} + D$.
\end{itemize} 

The energy evaluation of different ansatze are performed as follows: with given variational parameters \{$q$, $a_i$, $\theta_i$, $\bm{r}_0$\}, we first initialize spin configuration $\bm{S}_{\bm{r}}$ on a finite $192\times192$ triangular lattice (normalized on each site). Then we perform a Fourier transform to obtain $\bm{S}_{\bm{k}}$ and finally evaluate the energy in reciprocal space. Some notes about the variational parameters: the optimized $\bm{r}_0$ is typically off the lattice. 
The parameters \{$\theta_1$, $\theta_2$\} are found to be small, but not negligible especially for the $2k_F=|\bm{b}_1|/4$ case.

The optimized $Q$ indeed satisfies $Q=2k_F$ for any point in all the phase diagrams presented in this Letter. While the $192\times 192$ lattice may still have finite size effect, we can give an upper bound of the uncertainty: $\Delta Q < 2\pi / 192$. Furthermore, within our numerical accuracy ($\Delta H/H_\text{sat}^0 \lesssim 0.1 \%$), the phase boundaries obtained by both approaches (using super-cell, or using ansatze) are found to be the same.

We note that we applied the same variational calculation on the $192\times 192$ TL to the short range $J_1$-$J_2$ Heisenberg model~\cite{Okubo12,Leonov2015}. The results show a very slight variation of $Q$ as a function of magnetic field and anisotropy. In particular, for $J_1=-1$, $J_2=0.5$, the magnitude of the ordering wave number is $Q=|\bm{b}_1|/6$ in most regions of the $D-H$ phase diagram, except for the high-field region inside the SkX phase, where it deviates to $Q\approx |\bm{b}_1|/6.2$.

\section{ Hall Conductivity}

For completeness, we include the formulae that were used for computing the  Berry curvature and the Hall conductivity~\cite{Vanderbilt_book}.
The Berry curvature is defined as:
\begin{equation}
\Omega_n(\bm{k}) = -2 \,\text{Im} \,\langle \partial_x u_{n\bm{k}} | \partial_y u_{n\bm{k}} \rangle,
\end{equation}
where $| u_{n\bm{k}} \rangle$ is the cell-periodic Bloch function of the $n$-th band at momentum $\bm{k}$.

The Chern number of the $n$-th band is defined by integrating the Berry curvature over the first BZ:
\begin{equation}
C_n = \frac{1}{2\pi}\int_\text{BZ} d \bm{k} \Omega_{n}(\bm{k}).
\end{equation}

Finally, the Hall conductivity can be obtained by:
\begin{equation}
\sigma_{xy}= \frac{e^2}{\hbar} \sum_n \int_\text{BZ} \frac{d \bm{k}}{(2\pi)^2}  f\left( E_n(\bm{k}) - E_F\right) \Omega_n(\bm{k}),
\end{equation}
where $f\left( E_n(\bm{k}) - E_F\right)$ is the Fermi distribution function.

\end{document}